\documentclass[conference]{IEEEtran}
\usepackage{cite}

\ifCLASSINFOpdf
\else
\fi
\usepackage[cmex10]{amsmath}

\usepackage{algorithm}
\usepackage{algpseudocode}
\usepackage{amssymb}
\usepackage{pdfpages,amsthm}

\begin{document}
%
\title{An Adaptive Distributed Resampling Algorithm with Non-Proportional Allocation}
%
%
%

\author{\IEEEauthorblockN{\"Omer Demirel\IEEEauthorrefmark{1},
Ihor Smal\IEEEauthorrefmark{2},
Wiro Niessen\IEEEauthorrefmark{2}, 
Erik Meijering\IEEEauthorrefmark{2} and
Ivo F.~Sbalzarini\IEEEauthorrefmark{1}}
\IEEEauthorblockA{\IEEEauthorrefmark{1}MOSAIC Group, Center of Systems Biology Dresden (CSBD), \\
Max Planck Institute of Molecular Cell Biology and Genetics, \\
Pfotenhauerstr. 108, 01307 Dresden, Germany.
\\ Email: \{demirel,ivos\}@mpi-cbg.de}
\IEEEauthorblockA{\IEEEauthorrefmark{2}Biomedical Imaging Group Rotterdam, Erasmus MC, \\
University Medical Center Rotterdam, Rotterdam, The Netherlands. \\
Email: \{i.smal,w.niessen\}@erasmusmc.nl, meijering@imagescience.org}}
\maketitle

\begin{abstract}
The distributed resampling algorithm with proportional allocation (RNA)~\cite{Bolic2005} is key to implementing particle filtering applications on parallel computer systems. We extend the original work by Boli\'c \textit{et al}. by introducing an adaptive RNA (ARNA) algorithm, improving RNA by dynamically adjusting the particle-exchange ratio and randomizing the process ring topology. This improves the runtime performance of ARNA by about 9\% over RNA with 10\% particle exchange. ARNA also significantly improves the speed at which information is shared between processing elements, leading to about 20-fold faster convergence. The ARNA algorithm requires only a few modifications to the original RNA, and is hence easy to implement. 
\end{abstract}

\begin{IEEEkeywords}
Distributed resampling, particle filter, parallel computing, tracking, image processing.
\end{IEEEkeywords}

%
\IEEEpeerreviewmaketitle

\section{Introduction}
Particle filters (PF) have experienced impressive improvement since their introduction~\cite{Doucet2000,Doucet2001,Djuric2003} and are considered the \textit{de facto} standard tool to estimate and track targets with non-linear and/or non-Gaussian dynamics. Due to their computational cost, however, many PF applications are limited to small problems or require long execution times. In order to relax this issue by leveraging parallelism in modern hardware, Boli\'c \textit{et al}. introduced two distributed algorithms in their seminal work~\cite{Bolic2005}: the distributed resampling algorithm with proportional allocation (RPA) and one with non-proportional allocation (RNA). These algorithms enabled the development of PF applications that efficiently use modern multi-core and multi-processor hardware, such as computer clusters.   

Here, we propose a simple, yet effective improvement to RNA based on a \textit{randomized} particle-routing scheme with an \textit{adaptive} particle-exchange ratio. This \textit{adaptive RNA} (ARNA) algorithm improves the runtime performance and the efficiency of RNA. We benchmark these improvements in two situations of object tracking, where (1) the particles on all PEs are initialized at the location of the object to be tracked and with ground-truth velocity, hence testing the (\textit{tracking}) performance, and (2) the particles on only one PE are initialized near the object to be tracked, on all others they are initialized uniformly at random. The latter tests how fast information is shared between PEs once one of them converged to the object (\textit{information sharing}).


\section{Particle Filters}
\label{sec:sir} 
A generic PF algorithm consists of two parts: (i) sequential importance sampling (SIS) and (ii) resampling~\cite{Doucet2001}. A popular combined implementation of these two parts is the sequential importance resampling (SIR) algorithm~\cite{Doucet2001}. 

Recursive Bayesian importance sampling~\cite{Geweke1989} of an unobserved and discrete Markov process $\{\mathbf{x}_{k}\}_{k=1,\ldots ,K}$ is based on three components: (i) the measurement vector $\mathbf{Z}^k=\{\mathbf{z}_{1},\ldots ,\mathbf{z}_{k}\}$, (ii) the dynamics (i.e., state-transition) model, which is given by a probability distribution $p(\mathbf{x}_{k} | \mathbf{x}_{k-1})$, and (iii) the likelihood (i.e., observation model) $p(\mathbf{z}_{k} | \mathbf{x}_{k})$. Then, the state posterior $p(\mathbf{x}_{k} | \mathbf{Z}^{k})$ at time $k$ is recursively computed as:
\begin{equation}
\underset{\text{posterior}}{\underbrace{p(\mathbf{x}_{k} | \mathbf{Z}_{k})}} = \frac{\overset{\text{likelihood}} {\overbrace{ p(\mathbf{z}_{k} | \mathbf{x}_{k})}}\,\, \overset{\text{prior}} {\overbrace{p(\mathbf{x}_{k} | \mathbf{Z}^{k-1})}}}{\underset{\text{normalization}}{\underbrace{p(\mathbf{z}_{k} | \mathbf{Z}^{k-1})}}} \, ,
\end{equation}
where the prior is defined as:
\begin{equation}
p(\mathbf{x}_{k} | \mathbf{Z}^{k-1}) = \int p(\mathbf{x}_{k} | \mathbf{x}_{k-1}) \, p(\mathbf{x}_{k-1} | \mathbf{Z}^{k-1}) \, \mathrm{d}\mathbf{x}_{k-1}.
\end{equation}
PFs approximate the posterior at each time point $k$ by $N$ weighted samples (i.e., particles) $\{\mathbf{x}^i_k, w^i_k\}_{i=1,\ldots,N}$. This approximation is achieved by sampling a set of particles from an importance function (proposal) $\pi(\cdot)$ and updating their weights according to the dynamics and observation models. This process is called sequential importance sampling (SIS)~\cite{Doucet2001}. However, SIS suffers from  \textit{weight degeneracy}, whereby small particle weights become successively smaller and do not contribute to the posterior any more. To overcome this problem, a \textit{resampling} step is performed~\cite{Doucet2001} whenever the number of particles with relatively high weights falls below a specified threshold. In order to parallelize the SIR algorithm, one only needs to focus on the \textit{resampling} step, since all other parts of the SIR algorithm are local and can trivially be executed in parallel. The complete SIR algorithm is given in Algorithm~\ref{alg:sir}. 

\section{Classical RNA}
In a distributed-memory computer system with $M$ processing elements (PEs, $m=1,\ldots ,M$), the resampling step in RNA is performed locally by each PE. While the number of particles per PE hence remains constant, ensuring perfect data-balance (i.e., all PEs hold the same amount of data), the weight distribution across PEs can become unbalanced. This requires particle routing (i.e., dynamic load balancing (DLB)) in which every PE moves a constant fraction of its particles to another PE, such that the particle weights become more evenly mixed. A pseudocode for ARNA is shown in Algorithm~\ref{alg:rna}.

\begin{algorithm}
\caption{Sequential Importance Resampling (SIR)} 
\begin{algorithmic}[1]
	\State (P) Propagate all particles according to the transition prior: $\mathbf{x}_k^{(i)}\sim p(\mathbf{x}|\mathbf{x}_{k-1}^{(i)})$, $i=\{1,\ldots ,N\}$
    	\State (U) Update the weights taking into account the measurements at time $k$, $\mathbf{z}_k$, as $\tilde{w}^{(i)}_{k}=p(\mathbf{z}_{k}|\mathbf{x}_{k}^{(i)})w^{i)}_{k-1}$
	\State Renormalize the weights as $w^{(i)}_{k}=\tilde{w}^{(i)}_{k}/\sum_{j=1}^N \tilde{w}^{(j)}_{k}$
	\State Compute the estimate $\mathbf{\hat{x}}_k=\sum_{i=1}^N w^{(i)}_{k}\mathbf{x}_{k}^{(i)}$ 
	\State Compute $N_{\textrm{eff}} = (\sum_{i=1}^N (w^{(i)}_{k})^2)^{(-1)}$
	\State Resample if $N_{\textrm{eff}}<N_{\textrm{thresh}}$ using Systematic Resampling 
\end{algorithmic}
\label{alg:sir}
\end{algorithm}

\begin{algorithm}
\caption{Resampling with Non-proportional Allocation (RNA)} \label{alg:rna}
\begin{algorithmic}[1]
	\State Exchange $N_{\textrm{ex}}$ of particles with neighboring PEs
	\State Renormalize weights as $w^{(m, i)}_{k-1}=w^{(m, i)}_{k-1}/W_{k-1}$
	\State Perform (P) and (U) steps of SIR to get $\mathbf{s}^m_k$
	\State Compute the estimate $\mathbf{\hat{x}}_k^m$ and the sum of unnormalized weights $W_k^{(m)}$
	\State Resample $\mathbf{s}^m_k$ using the locally normalized weights $\tilde{w}^{(m, i)}_{k}=w^{(m, i)}_{k}/ W_{k}^{(m)}$
	\State Set the $i$-th weight to $w^{(m, i)}_{k}=W^{(m)}_k$ 
	\State Send $\mathbf{\hat{x}}_k^m$ and $W_k^{(m)}$ to the master PE
	\State The master PE computes $\mathbf{\hat{x}}_k$ and $W_k$ and broadcasts the result to all PEs
\end{algorithmic}
\end{algorithm}

\subsection{Particle Routing via Local Exchange}


The \textit{local exchange} method uses a fixed number of $N_p=N/M$ particles on each PE and also fixes the number $N_{\textrm{ex}}$ of particles to be exchanged. In this RNA configuration, the PEs are arranged in a ring topology and each PE sends $N_{\textrm{ex}}$ particles to its (counter-)clockwise neighbor in the ring. Since each PE only communicates with its neighbor, several rounds of communications are required until the weights are approximately evenly distributed and the accuracy of the particle representation of the posterior $p(\mathbf{x}_{k} | \mathbf{Z}^{k})$ is recovered. 

\subsection{Deterministic Particle Routing Schedule}
The \textit{local exchange} method with a particle-exchange ratio of 10\% or 50\% is a popular choice when implementing RNA~\cite{Bolic2005,Miguez2007,zenker2010parallel}. This avoids the need for application-dependent DLB schedules. Fixing $N_{\textrm{ex}}$ in the local exchange method, the DLB scheme is easier and faster to design and implement. However, since this DLB scheme is static, it does not adapt to the dynamics of the application, where different load imbalance situations may arise. 

\subsection{Ring topology}
In the original RNA, the PEs are arranged in a ring and only communicate with their adjacent neighbors. PE $P_m$ randomly selects $N_{\textrm{ex}}$ (out of its $N_p$) particles and sends them to PE $P_{m+1}$. Concurrently, it receives $N_{\textrm{ex}}$ new particles from $P_{m-1}$. While the ring topology leads to a simple communication schedule, it also has the lowest \textit{conductance} (i.e., speed of information spreading) from a graph-theory point of view. Thus, the information of ``good'' particle weights is shared only slowly across PEs. Furthermore, the performance of this DLB scheme in the ring topology degenerates as the number of PEs increases~\cite{Ghosh1996}. 

\section{Adaptive RNA}
We propose the adaptive RNA (ARNA) algorithm, which improves over the classical RNA by using dynamically adaptive particle-exchange ratios and randomized ring topologies. 

\subsection{Adaptive Particle-Exchange Ratio}
The traditional RNA uses a fixed particle exchange ratio that need to be set by the user. We relax this constraint by making $N_{\textrm{ex}}/N_p$ dynamically adaptive, allowing it to vary between $0\ldots 50\%$ as:
\begin{equation}
\label{eq:Nex}
N_\text{ex} = N_p \left[ 0.5 - \frac{0.5 (PE_\textrm{eff}-1)}{M-1} \right] \, . 
\end{equation}
Hence, $N_{\textrm{ex}}$ is negatively correlated with the tracking efficiency $PE_\textrm{eff}$, which is defined as:
\begin{equation}
PE_{\textrm{eff}} = \frac{ \left (\sum_{m=1}^M\sum_{i=1}^N w^{(m,i)}_{k} \right )^2} { \sum_{m=1}^M\sum_{i=1}^N (w^{(m,i)}_{k})^2 } \, , 
\end{equation}
where $w^{(m,i)}_{k}$ is the weight of $i$-th particle on $m$-th PE. $PE_{\textrm{eff}}$ measures the percentage of PEs that have already located the object and track it successfully.

The adaptive exchange rate in ARNA frees the user of fixing this parameter, and helps reduce communication-network congestion and thus increases the parallel performance. The advantage of this adaptive approach becomes more pronounced for high tracking accuracies, i.e., in the \textit{tracking} case.

\subsection{Randomized Ring Topology}
In a complete graph, information can be shared between any two PEs in single communication step. However, such all-to-all communication limits the parallel scalability of the algorithm. We introduce an improved (in the sense of faster mixing) DLB scheme for ARNA that has the same communication cost as the original RNA, i.e., the same number of send and receive operations per PE.

We exploit the power of \textit{randomization} methods, which are well-established for approximately solving NP-complete problems, such as the present one. 
As a simple change to RNA, we randomize the vertex labeling in the ring topology. This is equivalent to having a complete graph and selecting different, random Hamiltonian paths (i.e., paths that visit each node exactly once) in this graph. Projecting the complete graph onto a ring topology via a Hamiltonian path, each PE only communicates with two other PEs, as in the classical RNA. We use Fisher-Yates shuffling~\cite{fisher1949statistical} to efficiently compute randomized ring topologies. One could also apply other regular graphs with low maximum degree, but such topologies would require knowledge about the hardware network connecting the PEs in the actual machine. With no prior knowledge about process-to-PE assignment and hardware network topology, the present random ring labeling provides a simple tool to increase the efficiency of information spread in ARNA.

\subsection{Algorithm}
ARNA only requires a few minor modifications to RNA in steps 1 and 2. A pseudocode for ARNA is given in Algorithm~\ref{alg:arna}.
\begin{algorithm}
\caption{Adaptive RNA (ARNA)} \label{alg:arna}
\begin{algorithmic}[1]
	\State Randomize the PE topology using Fisher-Yates shuffle~\cite{fisher1949statistical}
	\State Update the particle-exchange ratio $N_{\textrm{ex}}/N_p$ according to Eq.~\ref{eq:Nex}. This requires a global communication in order to compute $PE_{\textrm{eff}}$.
	\State Exchange $N_{\textrm{ex}}$ of particles with neighboring PEs
	\State Renormalize weights as $w^{(m, i)}_{k-1}=w^{(m, i)}_{k-1}/W_{k-1}$
	\State Perform (P) and (U) steps of SIR to get $\mathbf{s}^m_k$
	\State Compute the estimate $\mathbf{\hat{x}}_k^m$, and the sum of unnormalized weights $W_k^{(m)}$
	\State Resample $\mathbf{s}^m_k$ using the locally normalized weights $\tilde{w}^{(m, i)}_{k}=w^{(m, i)}_{k}/W_{k}^{(m)}$
	\State Set the $i$-th weight to $w^{(m, i)}_{k}=W^{(m)}_k$ 
	\State Send $\mathbf{\hat{x}}_k^m$ and $W_k^{(m)}$ to the master PE
	\State The master PE computes $\mathbf{\hat{x}}_k$ and $W_k$ and broadcasts the result to all PEs
\end{algorithmic}
\end{algorithm}

\section{Benchmarks}
We benchmark the improvements of the proposed ARNA over RNA using an application from object tracking in fluorescence microscopy imaging~\cite{akhmanova2005,Komarova2009}. The goal here is to track the motion of small structures that are labeled with fluorescent dyes. From this, one cn then characterize the dynamics of those objects and quantify, e.g., their velocity, spatial distribution~\cite{helmuth2010beyond}, motion correlations, etc. 

\begin{figure}[]
\centering
\includegraphics[width=\columnwidth]{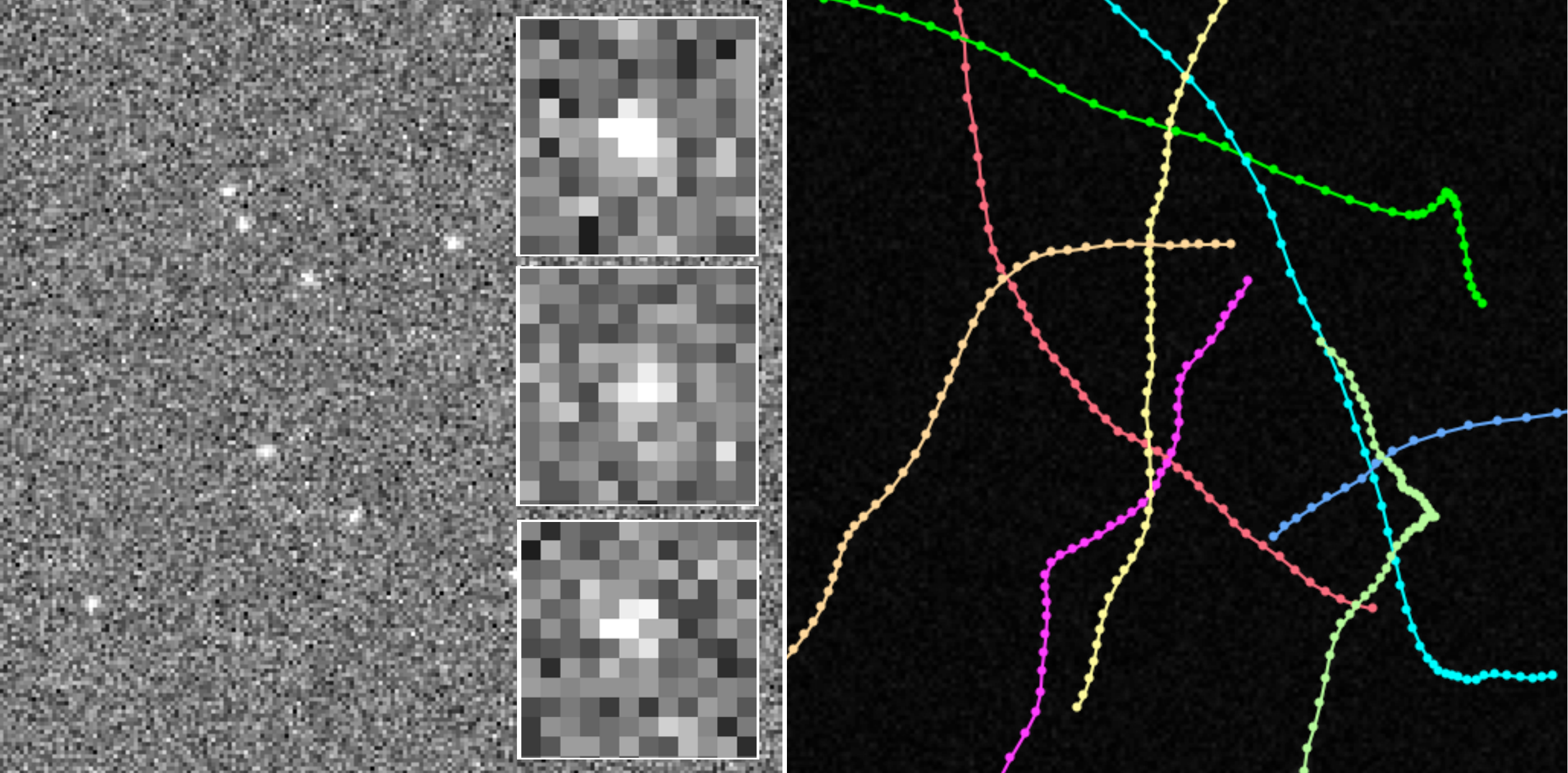}
\caption{Examples of synthetic images used in the benchmarks. Left: One frame of a typical 2D image sequence with signal-to-noise ratio 2, containing the small, bright objects of interest. Zoomed insets show noisy object appearance, modeled using a 2D Gaussian intensity profile corrupted with Poisson noise. Right: Typical object trajectories, generated according to the nearly-constant-velocity model.}
\label{fig_imagedata}
\end{figure}

We use the same previous sequential implementation of SIR~\cite{smaltmi,smal_media} inside both RNA and ARNA. The dynamics model assumes nearly constant velocity, and the appearance model approximates each object by Gaussian intensity profile in the final microscopy image. These are standard models that adequately describe biological fluorescence microscopy~\cite{smaltmi,smal_media}. The state vector in this case is $\mathbf{x}=(\hat{x}, \hat{y}, v_x, v_y, I_0)^T$, where $\hat{x}$ and $\hat{y}$ are the estimated $x$- and $y$-positions of the object, $(v_x,v_y)$ its velocity vector, and $I_0$ its estimated fluorescence intensity. An example image of how these bright objects then appear in the final, noisy microscopy images is shown in the left panel of Fig.~\ref{fig_imagedata}. The right panel of Fig.~\ref{fig_imagedata} shows some typical tracks along which these objects move during the time-course of a video.

For the performance evaluation, 10 different, synthetically generated image sequences are used, each containing 50 frames of size $512\times 512$ pixels. The tracking performance is evaluated for two different modes: \textit{tracking} and \textit{information sharing}. In the first mode, all PEs contain particles that are initialized at the true object state. In the second scenario, the particles are uniformly randomly initialized in state space on all but one PE. On one PE, the particles are initialized at the true state. This models the situation that one PE has discovered and converged on the object and needs to efficiently share this information with the other PEs. After that, the two distributed SIR implementations (one with ARNA and one with RNA) are used to locate the object in the subsequent frames and continue with accurate tracking and position estimation.

We compare ARNA against RNA with 0\%, 10\%, and 50\% particle-exchange ratios. 
The memory footprint of a single particle is 52\,kB (i.e., six doubles and one integer. The six doubles are the five components of the state vector and the particle weight. The integer is the process ID of where that particle belongs). All tests of \textit{tracking} are repeated 50 times for statistical significance. For \textit{information sharing}, we benchmark the recovery curve of  $PE_{\textrm{eff}}$ on five different synthetic image sequences, each test repeated 10 times. All experiments are run on the MadMax computer cluster of MPI-CBG, Dresden, which is equipped with 128\,GB DDR3 800-MHz memory per node and two Intel\textregistered \,  Xeon\textregistered \, E5-2640   six-core processors per node with a clock speed of 2.5\,GHz. Both ARNA and RNA are implemented in Java (v. 1.7.0\_13) in the Parallel Particle Filtering (PPF) library~\cite{demirel2013ppf}. We use OpenMPI's Java bindings (v.~1.9a1r28750) for inter-process communication, which are available as snapshot tarballs from the OpenMPI website~\cite{OpenMPITrunk}.

\subsection{Tracking performance}
\begin{figure}[]
\centering
\includegraphics[width=0.49\columnwidth]{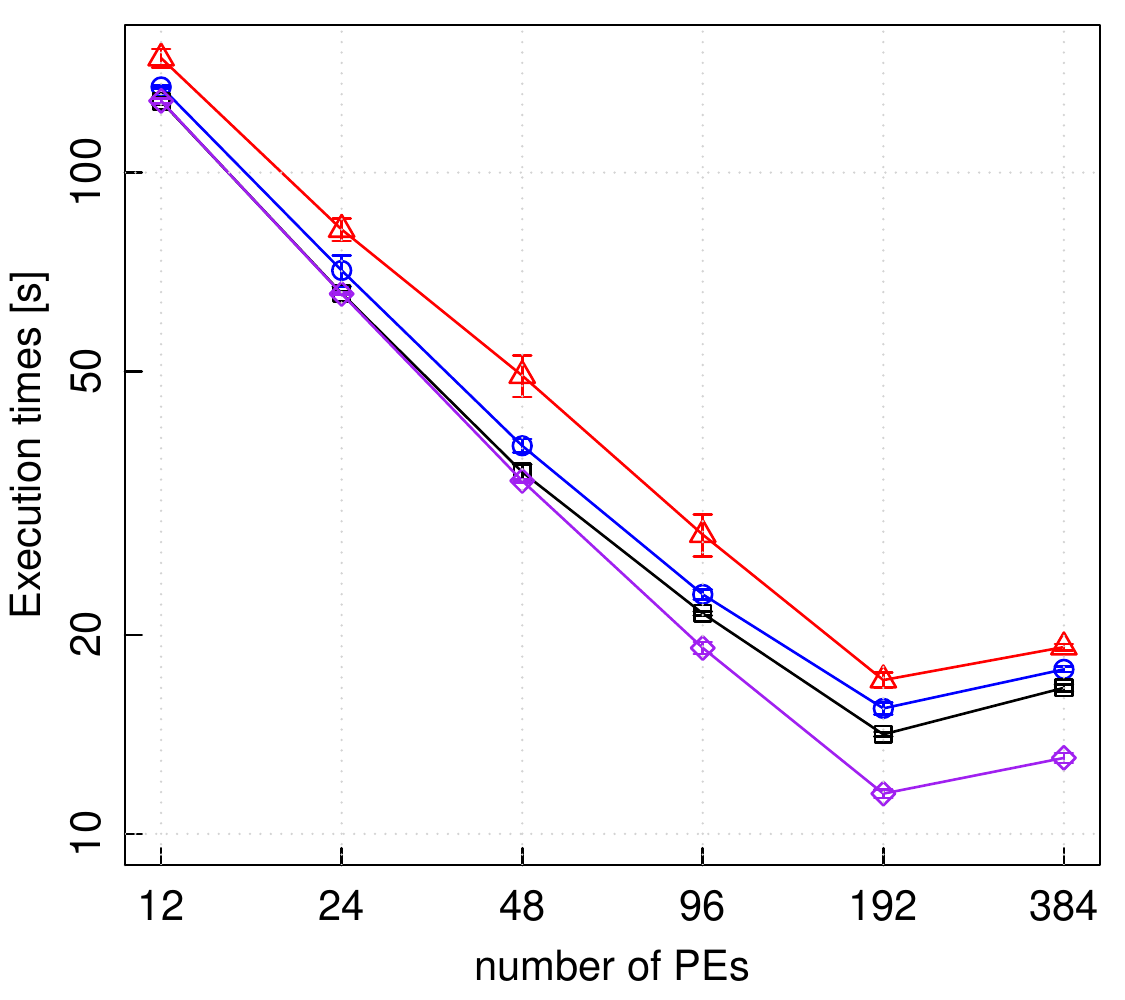}
\includegraphics[width=0.49\columnwidth]{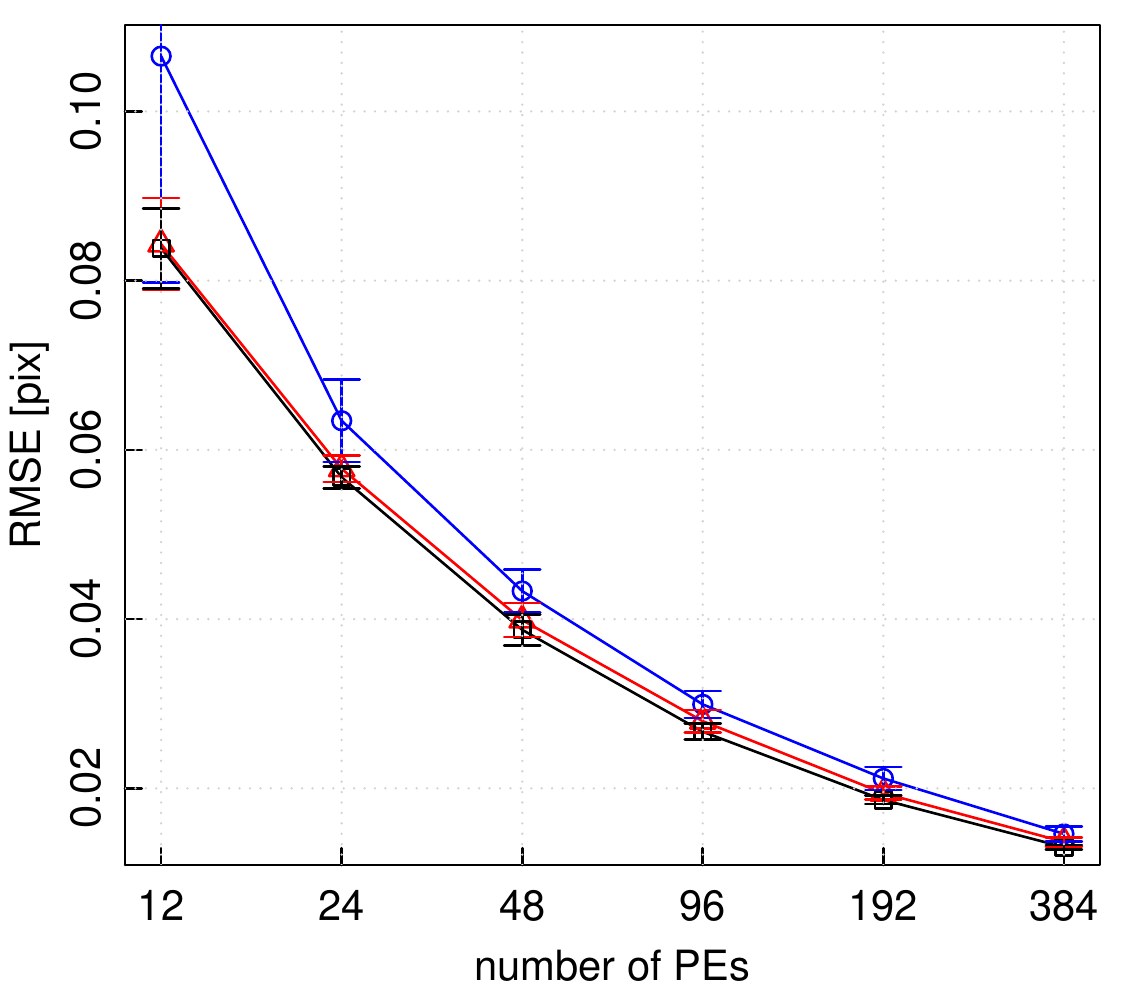}
\caption{Left: Execution times of RNA with 50\% exchange (red triangles), 10\% exchange (blue circles), and 0\% exchange (purple diamonds) compared with the timings for ARNA (black squares). A fixed total number of 19.2 million particles is distributed over an increasing number of PEs (strong scaling). ARNA is faster than RNA with 10\% and 50\% exchange. RNA with 0\% exchange (i.e., embarrassingly parallel RNA) defines the  lower bound for this test case, where no communication is necessary. 
Beyond 192 PEs, the number of particles per processor is too small to amortize the constant communication overhead. 
Right: RMSE tracking accuracy in pixels when using a constant number of 40 particles per PE, initialized at the target. 
RNA with 50\% particle exchange (red line) and ARNA (black line) show comparable tracking accuracy, whereas RNA with 10\% exchange (blue line) yields lower accuracy. As the total number of particles increases, the tracking becomes more accurate in all cases.}
\label{fig:strong_scaling_arna}
\end{figure}

We initialize 19.2 million particles at the location of the targeted object and thus we ensure high-accuracy tracking. In such a scenario, if correct dynamics and observation models are used, inter-process communication is virtually unnecessary since all PEs independently track the object. The classical RNA model, however, is oblivious to the mode of the application, as the process topology and the particle-exchange ratio are fixed. In ARNA the particle exchange ratio $N_{\textrm{ex}}/N_p$ is negatively correlated with the tracking efficiency. PEs do not exchange any particles if $PE_{\textrm{eff}}$ is above 99\%. The runtime results of the benchmarks are shown in Fig.~\ref{fig:strong_scaling_arna}. The tracking accuracy of ARNA is comparable to that of RNA with 50\% particle exchange. When exchanging only 10\% of the particles in RNA, the accuracy drops. Visually, however, all resulting trajectories are indistinguishable, as the Root Mean Square Error (RMSE) of the tracking is below 0.1 pixel in all cases.  

\subsection{Information Sharing Performance}
In applications with no prior information about the initial state of the system, it is common practice to initialize the particles uniformly at random 
throughout the state space. This helps explore the state space and first detect the object to be tracked. At some point, one of the PEs will (stochastically) detect the object to be tracked and the particles on the PE converge around the object. Until this point, all PEs uniformly sample the state space and communication between them does not help. Once one PE has found the target, however, this information should be disseminated among all PEs as quickly as possible, in order to allow the other PEs to contribute to the tracking accuracy.  In a parallel PF application we want all PEs to contribute to the result (i.e., not waste computational resources). $PE_{\textrm{eff}}$ should hence reach 100\% as quickly as possible after initialization. 

\begin{figure}[]
\centering
\includegraphics[width=0.49\columnwidth]{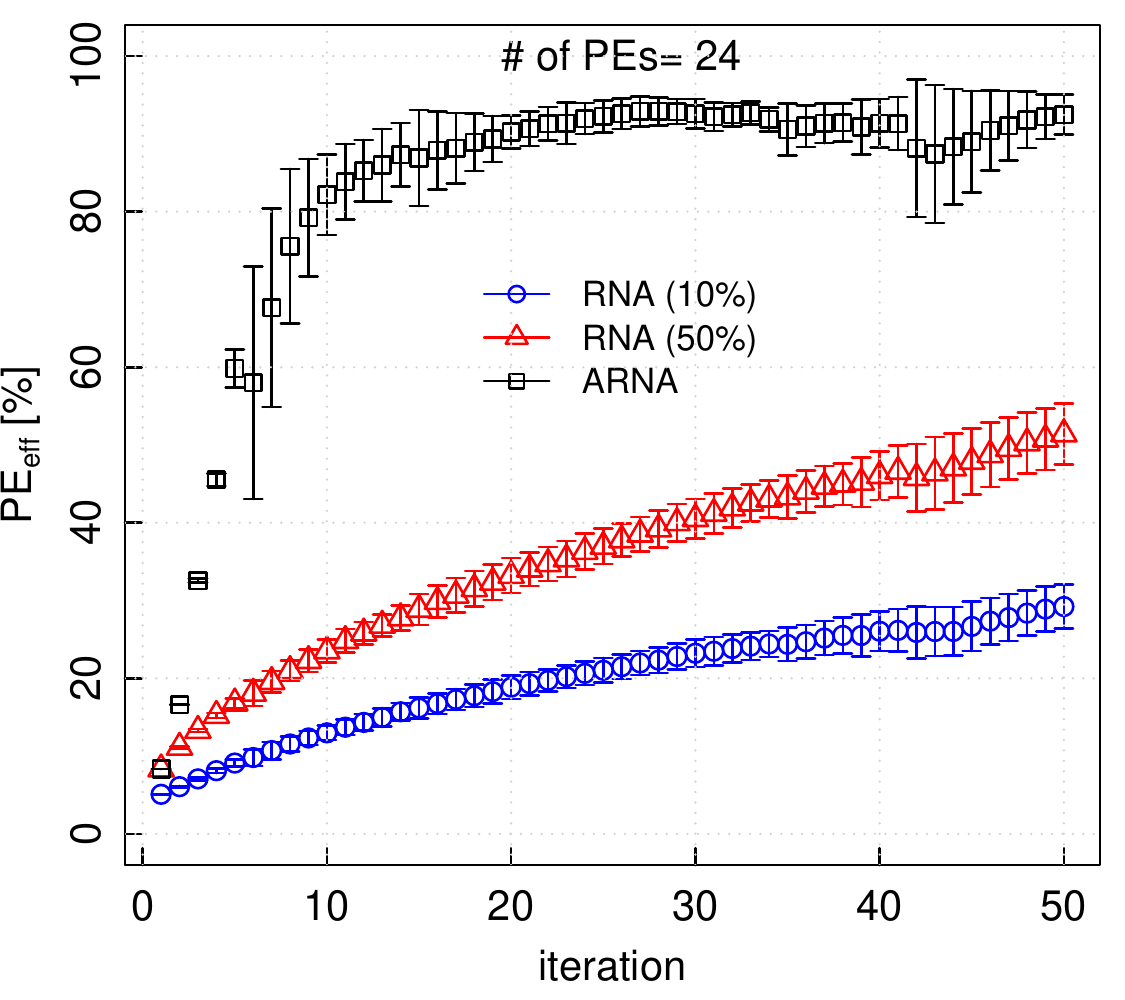}
\includegraphics[width=0.49\columnwidth]{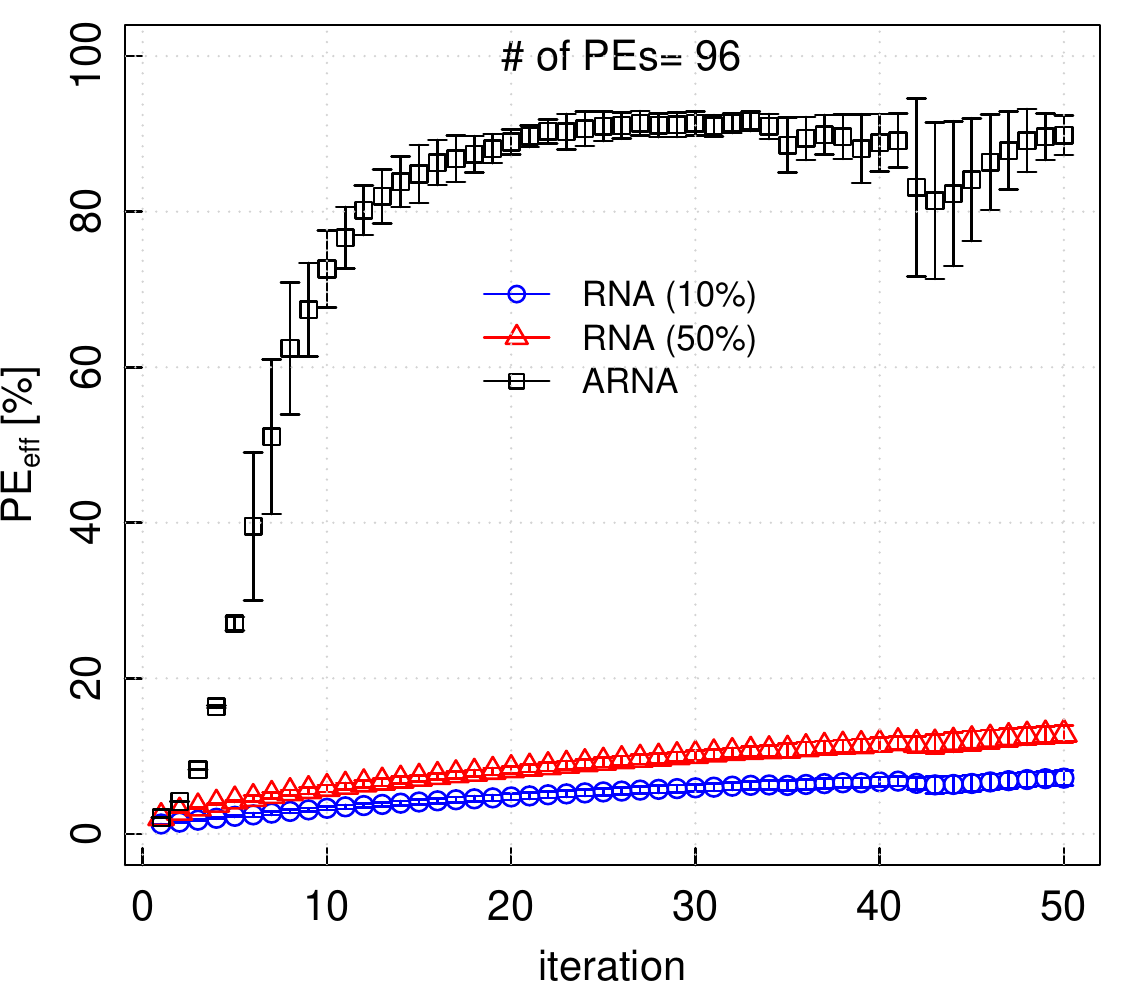}
\includegraphics[width=0.49\columnwidth]{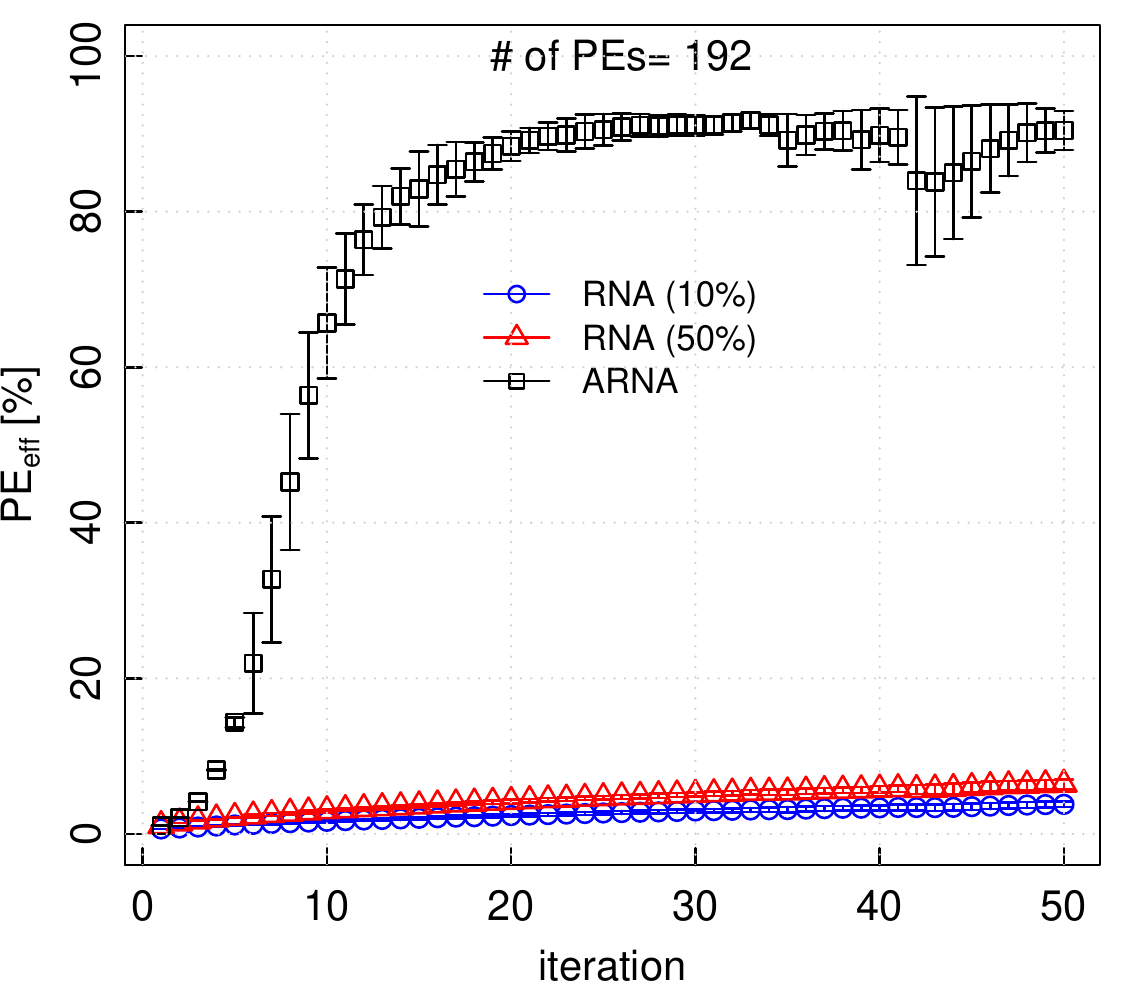}
\includegraphics[width=0.49\columnwidth]{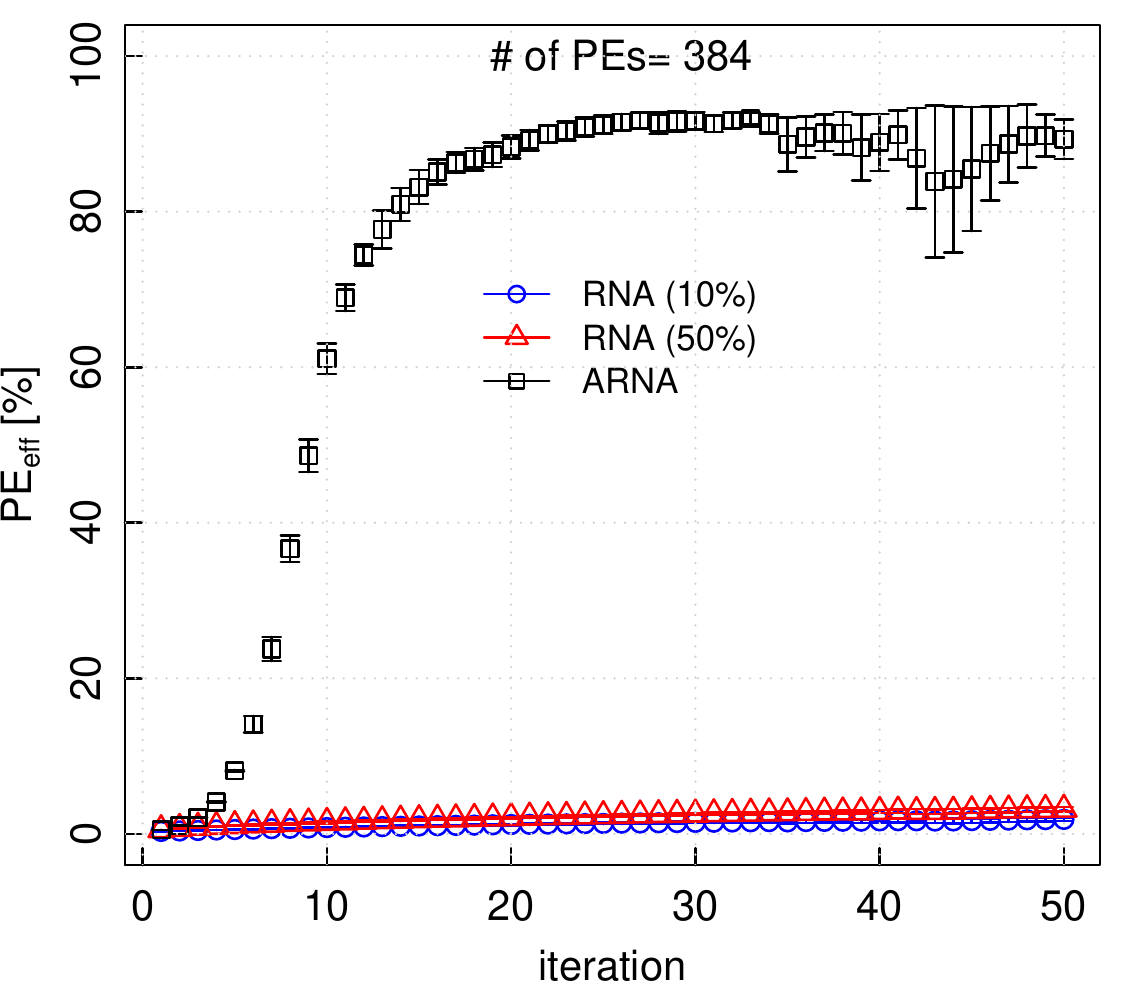}
\caption{Percentage of PEs engaged in successful tracking of the target ($PE_{\textrm{eff}}$) as a function of iteration number during the information sharing phase: ARNA (black), RNA with 10\% particle exchange (blue), and RNA with 50\% particle exchange (red) on 24 PEs (upper left), 96 PEs (upper right), 192 PEs (lower left), and 384 PEs (lower right). 
The randomized ring topology of ARNA leads to a faster spread of information and hence a higher proportion of computational resources that are engaged in contributing to problem solution.}
\label{fig:arna_recovery}
\end{figure}

In ARNA, the randomized ring topology helps share the detection information more rapidly. 
Figure~\ref{fig:arna_recovery} shows how $PE_{\textrm{eff}}$ evolves with algorithm iterations for the different parallel algorithms, counting iterations from the time point where one of the PEs has found the object. 

\section{Conclusions}
We presented ARNA, an adaptive randomized version of the classical RNA~\cite{Bolic2005} algorithm for parallel particle filtering (PF). ARNA uses a dynamically adapted particle-exchange ratio, which depends on tracking accuracy. This reduces redundant communication once the target is being successfully tracked. In such cases, only little communication is required, and ARNA is about 9\% faster than RNA with a 10\% exchange ratio. 
Randomizing the ring topology of the PEs changes the communication partners in each iteration, hence enhancing information sharing. This leads to a faster increase in the percentage of PEs that have successfully located the target once at least one PE has converged. ARNA hence improves the tracking accuracy and effectiveness by having more PEs contribute to the result earlier.

The tracking accuracy of ARNA is comparable with that or RNA with large exchange ratios. However, the fraction of PEs that contribute to this accuracy (i.e., tracking effectiveness) increases faster in ARNA than in RNA. In a network of 24 PEs, for example, $PE_{\textrm{eff}}$ of a 50\%-RNA is about 25\% after 10 communication rounds (iterations). When exchanging only 10\% of the particles, the fraction is only at 15\%. In ARNA, $PE_{\textrm{eff}}$ is larger than 80\% in the same situation. In a large network of 384 PEs, the difference between RNA and ARNA is even more pronounced: both RNA versions score below 4\% $PE_{\textrm{eff}}$, whereas ARNA reaches 60\% after 10 iterations, converging to over 80\% after 20 iterations. 

ARNA is easy to implement and requires only few minor changes with respect to RNA. Future work could further improve ARNA by including prior knowledge about how the processes are assigned to PEs and how the latter are connected in the machine by the hardware network. This would help optimize the ring topology such that neighboring PEs in the ring reside on the same cluster node, hence further reducing communication overhead. Using hardware-topology information would also enable the use of other regular graphs with low maximum degree as communication topologies, which may better reflect a specific hardware than a generic ring topology. 

The ARNA algorithm is implemented in Java as open source in the Parallel Particle Filtering (PPF) library~\cite{demirel2013ppf}, which is freely available for download from the MOSAIC Group web page at {\texttt{mosaic.mpi-cbg.de}}.


%



\section*{Acknowledgments}
The authors would like to thank the MOSAIC Group (MPI-CBG, Dresden) for fruitful discussions and the MadMax cluster team (MPI-CBG, Dresden) for operational support. \"{O}mer Demirel was funded by grant \#200021--132064 from the Swiss National Science Foundation (SNSF), awarded to I.F.S. Ihor Smal was funded by a VENI grant (\#639.021.128) from the Netherlands Organization for Scientific Research (NWO).

\ifCLASSOPTIONcaptionsoff
  \newpage
\fi



%
\bibliographystyle{unsrt}
\bibliography{particle_filtering}

%

%
%



\begin{IEEEbiography}[{\includegraphics[width=1in,height=1.25in,clip,keepaspectratio]{portrait_Ivo_SBALZARINI.jpg}}]{Ivo F.~Sbalzarini} is the founder and head of the MOSAIC Group. He received his Diploma (equiv.~M.Sc.) in Mechanical Engineering, with majors in Control Theory and Applied Mathematics, from ETH Zurich (Switzerland) in 2002, which was awarded the Willi Studer Prize for the best Diploma in Mechanical Engineering of that year. 

In 2006 he obtained his Ph.D.~in Computer Science from ETH Zurich under the supervision of Professor Petros Koumoutsakos. His Ph.D.~thesis was awarded the Dimitri N.~Chorafas Research Award for 2006.
From 2006 until 2012, he was Assistant Professor of Computational Science at ETH Zurich. He was also invited Professor of Biology at \'{E}cole Normale Sup\'{e}rieure in Paris (France) and Research Group Leader in Bioinformatics at the Mediterranean Institute for Life Sciences (MedILS) in Split (Croatia). Since 2012 he is a Senior Research Group Leader with the Center of Systems Biology Dresden at the Max Planck Institute of Molecular Cell Biology and Genetics (MPI-CBG) in Dresden (Germany). 

He is a life-long honorary member of the Technical Society of Zurich, a path co-leader of the Center for Advancing Electronics Dresden, an Associate Editor for BMC Bioinformatics, a member of the Program Committee of the IEEE Symposium on Biomedical Imaging (ISBI) since 2010, and serves on various conference, program, and reviewer boards. His research interests include image analysis and image processing, adaptive discretization schemes for PDEs, randomized algorithms, and parallel algorithms.
\end{IEEEbiography}




\end{document}